%% file: main.tex
\documentclass[conference,10pt]{IEEEtran}
\usepackage{algorithm}
\usepackage[bf,SL,BF]{subfigure}
\usepackage{algorithmicx}
\usepackage{algpseudocode}
\usepackage{amsmath}
\usepackage{graphicx}
\usepackage{subfigure}
\title{External Memory based Distributed Generation of Massive Scale Social Networks on Small Clusters}

\author{\IEEEauthorblockN{Sandeep Gupta}
\IEEEauthorblockA{San Diego Supercomputer Center\\
University Of California, San Diegoy\\
San Diego, Californai, 92093\\
Email: sandeep@sdsc.edu}
}

\begin{document}
\maketitle
\begin{abstract}
Small distributed systems are limited by their main memory to generate massively large graphs. Trivial extension 
to current graph generators to utilize external memory leads to large amount of random I/O hence do not scale with size. 
In this work we offer a technique to generate massive scale graphs on small cluster of compute nodes with limited main memory. We develop several distributed 
and external memory algorithms, primarily, shuffle, relabel, redistribute, 
and, compressed-sparse-row (csr) convert. The algorithms are implemented in MPI/pthread model to help
parallelize the operations across multicores within each core. 
Using our scheme it is feasible to generate a graph of size $2^{38}$ nodes (scale 38) using only 64 compute nodes. This can be compared 
with the current scheme would require at least 8192 compute node, assuming 64GB of main memory. 

Our work has broader implications for external memory graph libraries such as STXXL and graph processing on SSD-based supercomputers 
such as Dash and Gordon~\cite{Norman:2010:ADS:1838574.1838588,strande:2012:GDP:2335755.2335789}.

\end{abstract}

\section{Introduction}
\label{sec:intro}
In the last decade social networks surfaced on WWW and became immensely popular. With the advent of these graphs,
massively parallel breadth first search (BFS) over scale free graphs has become an active area of study.
Researchers have studied the problem under different combinations of hardware, runtime, and parallel programming model.
In order to compare and contrast the various approaches, the supercomputing community initiated a Graph500 challenge
recently. The challenge ranks the data intensive computing capability of supercomputers.
The current data deluge has evidently marked a shift in 
the supercomputing needs from traditional numerical computation to data-intensive computing. 

This work focuses on scalable implementations of the first of the two Graph500 kernels, namely, the graph generation kernel. 
There, the scheme for generating graphs utilizes the R-MAT model~\cite{DBLP:conf/sdm/ChakrabartiZF04}. R-MAT model is invoked 
by each core to generate random edges, creating a total of $f *n$ edges, where f is edge factor and n is the number of nodes desired in the output graph.
The degree distribution of the graph thus generated is biased with respect to the node identifier, identifiers with smaller 
values have high degree distributions. Distribution and further processing of 
this graph can lead to locality both in local memory and across compute nodes that may reduce the validity of any operation 
or hinder performance. Therefore, the graph must be ``de-biased''. This is achieved by post processing the graph using 
an MRG hash function. 
The suggested MRG hash function is fast, has no collisions, and, produces ``high quality'' permutations. These facts make
it a good choice for main memory based graph generation.
The entire scheme is main memory based: all of the graph is stored in the main memory and all the processing also happens from 
the main memory. We refer to this kernel as the `hashing based' kernel in the rest of this paper.

The kernel requires 8 terabytes of main memory to generate a graph of size $2^{34}$ (scale 34) nodes. 
For `large' graphs (scale 34 and above) the dataset size is even larger, doubling at every scale.  
To meet such  memory demands clusters of size upwards of $4096$ compute nodes, assuming $64GB$
memory per compute node, would be required.
This has limited the generation of large graphs to select few largest supercomputers in the world, and that too,
often by reserving the entire supercomputer.
Smaller systems invariably fall short on both the memory and the time to generate competitive size graphs using this scheme. 

One possible way to alleviate the memory limitation issue is to use external memory~\cite{He:2010:DRF:1884643.1884661,5644845}.
The kernel can be extended trivially for generating graphs on external memory.
However, the R-MAT model generates edges in random order. Hashing based method processes
these edges in the same order as they are generated. Now, if the generated edges are stored on disk,
this ordering will incur heavy amounts of random I/Os. Hence, a naive extension to graph generation over external
memory would be inefficient, and infeasible for very large graphs. Caching and buffering cannot help because the main memory buffers are 
extremely small compared to total storage. 

In this work, we offer a technique to generate massive scale graphs on small cluster of compute nodes (approximately 64 compute nodes). We develop several distributed 
and external memory algorithms, primarily, shuffle, relabel, redistribute, 
and, compressed-sparse-row (csr) convert. For each of these, we present a distributed, hybrid, MPI/pthread based implementation to
help in parallelizing the operations across multicores within each compute node. We remark that it is possible to utilize 
multicores using MPI alone as well. However, the communication cost in that case is significantly higher. 
The relabeling algorithm resembles the sort-merge-join algorithm\cite{Silberschatz:2005:DSC:993519} popular in database domain.
Two approaches are presented for redistribution and csr. In the first one, the edges are treated unordered. In this approach,
building csr becomes expensive and we, therefore, present a parallel version.
In the second approach, the edges are processed in the order determined by relabeling phase; this simplifies and expedites 
the csr phase considerably. Amongst all these algorithms, relabeling turns out to be the most expensive, 
it's overall cost being slightly more than hashing. (On our machine, hashing $2^{30}$ integers required 1.34 seconds 
while sorting them into 65,536 sized chunks requires 5.134 seconds.) However, this cost is well compensated for
by the increased efficiency obtained as a result during the redistribution and the csr phases. 
Except shuffle, all the phases utilize external memory.

The proposed algorithms together are able to overcome the memory limitations of the hash based kernel.
They are able to generate same size graphs in much less memory (or generate much larger graphs using the same amount of memory).
For example, we can generate 16 billion node (2048 GB) graph using only (256 GB memory, or 4  compute nodes) while the present 
approach would require at least (2048 GB memory) or (32 GB) compute nodes. 
Our implementation has been able to generate graphs as large as $2^{37}$ nodes using $32$ compute nodes, each with $64 GB$ memory.

\input{approach_v2.tex}

\section{Conclusions and Future Work}
Generating large graphs is currently possible only on select few supercomputers due to the immensely high memory
requirement. 
The algorithms presented in this paper offer an alternative method
that can be used for generating large graphs on more affordable supercomputers. We use external memory. This facilitates 
reducing the burden on the main memory, helping in scaling the algorithm to much larger sizes than those possible through 
the hash based kernel. The use of external memory is also advantageous because it is much cheaper and offers lot more storage than main memory.
After this, the only true limitations for large graph generation are the number of available cores and the IOPS of the SSDs.

Experiments demonstrate good strong and weak scaling of our approach in presence of limited main memory. 
Further experiments could not be carried out due to limited time availability constraints. Based up the current performance 
and the observed linear scaling of the approach, we project that $2^{38}$ sized graph possible using only modest 64 compute nodes in a reasonable time duration (less than 6 hours) as compared
to upwards of 8192 compute nodes required with memory based approach.

A comparison of our approach with that of one implemented using map/reduce engine would be interesting. 
More so, because several operations in the graph generation phase can naturally be expressed using map and reduce operators. Such a comparison
would inform us about relative importance of using low latency interconnects, pipelining, and will bring out the advantages and disadvantages 
of map/reduce framework over the MPI/pthread framework. We are investigating this aspect and will address it in the follow up work. 

Finally, the sorted merge based distributed processing ideas developed here  have  
implications on external memory parallel graph libraries such as~\cite{Dementiev:2005:SST:2104638.2104695} and external memory graph processing algorithms~\cite{Ajwani:2009:DEE:1575424.1575426}.

\section{Acknowledgments}
I would like to thank the reviewers for their time and insightful comments. 
I thank John Feo for discussions external memory and distributed memory shuffle algorithm, Allan Snavely for relentlessly 
pursing massive scale graph processing using SSDs  and giving me the opportunity to work on this problem, and 
Chaitan Baru for discussions on labeling problem and it's connection with join operator in DB2.
I also thank Mahidhar Tatineni and Robert Sinkovits for helping with installation, running, and debugging issues on Trestles. 
The experiments were supported through XSEDE startup award for allocations on  several supercomputers.

\bibliographystyle{IEEEtran}
\bibliography{main_v6}

\end{document}

%% file: approach_v2.tex
\section{Preliminaries}
\label{sec:prelim}
We begin with describing a few preliminary notations to be used in this paper.
and also review briefly the state of art sequential implementation of R-MAT based scale free graph generator.
\begin{itemize}
\item Vertex: A node in the graph and represented as $u, v,w$ etc.
\item Edge:  An undirected  pair $(u,v)$
\item Adj(u):  List of vertices adjacent to $u$
\item Graph: G= (V,E) represents a graph with $n = |V|$ vertices 
and $m = |E|$ edges. 
\item CSR(G): compressed row representation csr of any graph $G$ consists of two vectors namely an offset vector (offv) and an adjacency vector (adjv). 
The offv indexes into the adjv vector. The adjacency (edge) information is stored in  adjv vector. The neighbors of node $nid$ are 
stored as entries in adjv vector from  range $[offv[nid], offv[nid+1]]$.
\item Storage cost $\mathcal S$: Gives the number bytes required to store a particular data object of its type. For example ${\mathcal S}(int)$ is 8 bytes in our setup.
\item Range partitioning:
Given objects (vertices) with identifiers  $[0:n]$ and value $k$, the range partitioning creates $k$ partitions
each of size $w = n/k$ such that partition $p$ has objects with identifiers in range $[w * i: w*(i+1)]$. We use $RP(n,k)$ to denote range 
partitioning into $k$ ranges. 
\end{itemize}

\paragraph*{Steps in graph generation}

The specification of the graph generator can broadly be described as consisting of 
four phases: 1) the permutation phase, 2) the edge generation phase, 3) the relabeling phase, and, 4) building the
CSR phase. It takes as input the edge factor $f$ and the number of graph nodes $n.$ 

The permutation phase produces $P,$ a vector of $n$ integers in range $[0:n]$ shuffled randomly. 
An edgelist consisting of $n*f$ edges is generated via call to a function $gen\_rmat\_edge()$~\cite{DBLP:conf/sdm/ChakrabartiZF04}. 
Edges are then relabeled either using the
permutation vector, i.e., an edge $e = (u,v)$ in the edgelist is relabeled as $(P[u], P[v]).$ 
Next, the edgelist is sorted using the source field. The sorted edge list is finally
used to build the CSR representation as described in algorithm~\ref{alg:buildcsr}. 

This is a sequential implementation of the generation. More details on it as well as a parallel distributed implementation
can be found on \cite{graph500} for the interested reader. In either case,
generating and utilizing the permutation vector typically turns out to be very expensive at large scale. It is possible to
by-pass/avoid permutation by using a perfect hashing function such as MRG~\cite{L'Ecuyer:1993:SGM:169702.169698}, as 
implemented in the Graph500 kernel~\cite{graph500} (and briefly discussed in the section~\ref{sec:intro}).

As discussed earlier, the hashing technique requires a lot of main memory making it infeasible to generate large graphs
on smaller clusters. Hence, our effort is to design a scheme that can be implemented on such clusters. It turns out 
that following the Graph500 specifications (utilizing the permutation vector etc.) is more beneficial to this end.

\begin{algorithm}
\caption{build\_csr(edgelist el)\label{alg:buildcsr}}
\begin{algorithmic}[1]
\Require edgelist $el$ be sorted 
\State aidx = 0, offv[0] = aidx, elidx = 0, csrc = 0
\While {1}
\If {$csrc \neq el[elidx].src$ }
\State ++csrc 
\State offv[csrc] = elidx
\State continue
\EndIf 
\State adjv[elidx] = el[elidx].des, ++elidx
\EndWhile
\end{algorithmic}
\end{algorithm}

\section{Hybrid MPI/pthread R-MAT generator}
We describe our implementation of the distributed R-MAT generator using the hybrid pthread/MPI framework. 

\subsection{ External Memory based Distributed (hybrid mode) R-Mat  Generator}
We begin with the description of our setup for generating graphs using the external memory. 
Our setup consists of $nb$ compute nodes each with  $nc$ cores. Following variables are used:
\begin{itemize}
\itemsep=0pt\topsep=0pt
\item Bucket size : B =  n/nb
\item Bin size : b = B/ nc
\item Compute node  rank : bid
\item Core rank : cid
\item Edge block capacity: $C_e.$  Number of edges per disk block.
\item Memory per core : $mmc$
\end{itemize}
We range partition the nodes across the compute nodes. Each range is further partitioned
and distributed across the cores. We say that the core is the ``owner'' of the nodes that belong 
in its partition range. A core also ``owns''  the edges whose source is from its partition range. 

We assume an abstract data type  edgelist that supports all the standard operations
such as insert, sort, and the like, \textit{except} only the delete operation. We have implemented an external memory 
data structure for the edgelist. 
Since there is partitioning across the cores, we need the notion of chunk partitioning in addition to range partitioning.
\paragraph*{Chunk partitioning}
Given a  collection $C$ (mostly edges) and  a chunk size $csz$, chunk partitioning is defined as decomposing 
the collection into $k = |C|/csz$ chunks, each of size $csz.$ We use $CP(C,csz)$ to denote chunk partitioning.

Furthermore, we also introduce what we call the `$k:1$ scatter-gather MPI/pthread communication pattern' framework. 
This is the framework that we will use for implementing most of our steps requiring communication.

\paragraph*{k:1 Scatter-gather Hybrid MPI/pthread communication pattern}
In this framework, each compute node has $k$ scatter threads and 1 gather thread. 
The scatter threads at any given compute node $i$ generates data for every other compute node $j$. 
The data at each compute node is collected by a gather thread which then further processes it. 


We now use the notations and ideas introduced to describe our parallel, scalable, external memory implementation of
the distributed R-MAT generator.

\subsection{The algorithms}
In designing our hybrid implementation of the R-MAT generator, we adhere to the same sequence of main steps described in section~\ref{sec:prelim}. In particular, 
the step by step procedure is: build a permutation of the nodes, then
generate edges, then relabel the edges, and finally, build the CSR format of the graph.
However, we redesign the algorithm for each step so that it is implemented  using
distributed processing, shared memory, parallel implementation, and the $k:1$ communication pattern described earlier.
The primary objective is scalability on smaller machines with external memory and fast communication network.

\subsubsection{The driver routine}
The graph generator is invoked through the main program that takes as input $n$ and $f$, the number of 
nodes and edge factor respectively. It then launches MPI processes on each compute node with same parameters.
Each MPI process executes the steps, in that order, and in sync with other processes. When necessary each MPI process 
also launches $nc$ pthreads, one per core, for example, during edge generation and relabeling steps. We now describe 
our implementation of each of the steps.

\subsubsection{Distributed Random Shuffle}
Algorithms~\ref{alg:send}, \ref{alg:recv}, and~\ref{alg:shuffle} describe the distributed shuffle algorithm
in the hybrid MPI/pthread environment. 

Algorithm~\ref{alg:shuffle} produces a permutation vector $pv$ of integers $[0:n].$
To do so, each compute node maitains two buffers $rbuf$ and $sbuf$. The buffer $sbuf$ is  initialized
to a unique partition obtained from range partitioning $RP(n, nb)$.  
In each step, the compute nodes shuffle the $sbuf$ buffer.
They then send portions of the $sbuf$  to other compute nodes with the help of threads~\ref{alg:send} 
and~\ref{alg:recv} using $1:1$ scatter gather pattern. The received messages are stored in $rbuf.$
The algorithm swaps the buffer at the end of the step. 

This shuffling and exchange is carried out for  ${\mathcal O}(\log n)$ steps.
Each step requires ${\mathcal O}(n/nb \log n/nb)$ and $O(nb^2)$ message transfers. 
The total time for shuffling, therefore, is ${\mathcal O}(n k^2\log^2 n )$ where $k = nb.$
Upon completion the resulting permutation vector $pv$ is distributed across the compute nodes. 
This distribution inherently 
induces a chunk partitioning of range $[0:n]$ with chunk size $csz$ equal to bucket size $B.$ (We remark that the chunk in this case is ordered.)
We use $pv_i$ to denote the $i^{th}$ range stored at compute node $i.$


\begin{algorithm}
\caption{send()\label{alg:send}}
\begin{algorithmic}[1]
\For{$j \in [0:nb]$}
\If {$j \neq bid$}
\State MPI\_Send(sbuf + j * b , b)
\EndIf
\EndFor
\end{algorithmic}
\end{algorithm}

\begin{algorithm}
\caption{recv()\label{alg:recv}}
\begin{algorithmic}[1]
\For{$j \in [0:nb]$}
\If {$j \neq bid$}
\State MPI\_Recv(rbuf + j * b , b)
\EndIf
\EndFor
\end{algorithmic}
\end{algorithm}

\begin{algorithm}
  \caption{distributed\_shuffle()\label{alg:shuffle}}
\begin{algorithmic}[1]
\State iter = 0
\Repeat
\State shuffle(sbuf)
\State launch send and recv threads
\State wait for send and recv threads
\State rbuf[bid * b, bid*b + i] = sbuf[bid*b, bid*b +i]
\State swap(sbuf, rbuf)
\Until{$iter < \log_{nb} n$}
\item MPI\_Barrier(MPI\_COMM\_WORLD)
\end{algorithmic}
\end{algorithm}

\subsubsection{Generate Edge list}
Our algorithm for edgelist generation is a direct extension of the sequential implementation to it's
parallel counterpart based on hybrid MPI/pthread framework.
Each compute node generates $n*f/nb = B *f$ edges. The load of edge generation is distributed further among the cores, i.e.,
each core is responsible for $B*f/nc = b * f$ edges. 
The collection of all edgelists across all the cores forms the distributed graph. 
The  I/O cost of this step is simply $O(b*f/C_e)$ sequential I/Os.
\begin{algorithm}
\caption{generate\_edgelist(tid)}
\begin{algorithmic}[1]
\item call append(el, gen\_rmat\_edge()) for $b * f$ number of times
\end{algorithmic}
\end{algorithm}

\subsubsection{Relabel Edges}
Edge generation is immediately followed by relabeling. This is the step that captures the central idea  of our approach. 
In this step, each core relabels the source and destination of the vertices
as per permutation vector, i.e., node with identifier $i$ is assigned a new label $pv[i].$  We first relabel the 
destination vertices of each edge and then the source vertices.
The algorithm is described for the destination field. Relabeling of source field works the same.

In order to relabel the edges efficiently, we perform chunk partitioning of $el$ 
with chunk size as $mmc$. Let $cl = CP(el,mmc)$ be the collection of chunks. The collection $cl$ is again chunk 
partitioned into $nc$  groups each of size $|cl|/nc,$ i.e., $gl = CP(cl, nc).$  Each group of chunks is then associated with a core.

The relabel algorithm, illustrated in~\ref{alg:label_edges} is executed by each core, 
which begins by  sorting (based up on destination field of the edge) all the chunks in its group,
one chunk at a time. 

The cores then work in a lock step fashion. At any time step $t,$ the cores perform 
a sort-merge-join style algorithm between $pv_t$ and destinations in range $rp_t$.  
The ordered permutation chunk $pv_t$ has the new identifiers (or labels) for the identifiers that lie in partition $rp_t.$
Since, the permutation range $pv_t$ is on remote server $t$, it needs to be fetched before the merge-join can take place.  
Core $0$ fetches the permute range using operation $get\_permute\_range.$  A $permute\_server$ thread (not depicted here) 
is launched (after the shuffle step) on each compute node that serves the permutation ranges to the requesting compute node.
Once the permutation range is in the local buffer, each core performs the sort-merge operation for each of 
the edgelist chunk in the group as depicted in algorithm~\ref{alg:label_edges} 
(line 12--17) and algorithm~\ref{alg:label_chunk}.

The algorithm is repeated again to relabel the source field.  The I/O complexity of this step is dictated by 
the number of disk blocks read by each core. In the current setup, this equals to 
$O(\frac {2* b*f * \mathcal S(int)} {C_e})$ sequential I/Os.

\begin{algorithm}
\caption{$label\_chunk(id, pid, elc, elci)$\label{alg:label_chunk}}
\begin{algorithmic}[1]
\While{elc[elci].des == id}
\State elc[elci].des = pid \Comment{Assign the new identifier and labels here}
\State ++elci
\EndWhile
\State return elci
\end{algorithmic}
\end{algorithm}

\begin{algorithm}
\caption{$label\_edges(tid)$\label{alg:label_edges}}
\begin{algorithmic}[1]
\For {each edgelist chunk $ec$ in group $gl[tid]$}
\State read into main memory \Comment{If chunks are stored on disk}
\State sort $ec$  \Comment{Either on source or destination field}
\State write sorted $ec$ on disk
\EndFor
\State $id = 0$
\For {$s \in [0:nb]$} 
\If {$tid == 0$}
\State $pv_s \longleftarrow get\_permute\_vec(s)$ \Comment{$pv_s$ is shared across}
\EndIf
\State wait(label\_wait\_barrier)
\For {$pv \in pv_s$}
\For {$i \in [0:|gl[tid]|]$} \Comment{process each chunk in its group} 
\State $elci_i \longleftarrow label\_chunk(id,pv, gl[tid][i], elci_i)$ \Comment{A sort merge join style operation}
\EndFor
\State $++id$
\EndFor
\State wait(relabel\_wait\_barrier)
\EndFor
\end{algorithmic}
\end{algorithm}

\subsubsection{Redistribute Edges}
\label{sec:redise}
Once the edges have been labeled they need to be shipped to their respective compute nodes.
In our setup an edge is owned by the same compute node that owns the source of the edge. 
Hence, if the (relabeled) source of an edge belong in range partition $rp_i,$ it is shipped to compute node $i.$

Our implementation of redistribution is an instantiation of the hybrid pthread/MPI $1:1$ scatter gather pattern as there are two 
threads per compute node involved to perform this task. The first thread performs the redistribution and
the second thread performs collection. They are conceptually similar to map and reduce operations in map/reduce framework but 
implemented using mpi blocking communication primitives.

The scatter thread executes the function $redistribute\_edges.$ depicted in algorithm~\ref{alg:redistribute_edges}. It iterates over the (relabeled) edges 
placing them in the corresponding outstanding packets (denoted as $elp_0, elp_1, \dots,$ in the algorithm description) up on the source of the edge.
If the buffer is full, it is sent to the collector thread (not depicted here)  which appends it to its local edgelist. 
The collected edgelist at compute node $i$ is ``owned'' by it, i.e., the source of the edges belongs to range $rp_i$
in the range partitioning $RP(n, nb).$

Assuming uniform distribution of edges across the compute nodes, the total I/O complexity of the redistribute step 
is $O(B*f/C_e).$ We, however, remark that a more efficient technique that deploys all the cores
with I/O complexity of $O(b*f/C_e)$ is feasible. 
\begin{algorithm}
\caption{redistribute\_edges()\label{alg:redistribute_edges}}
\begin{algorithmic}[1]
\For {$e \in el$}
\State $d \longleftarrow O(e.src)$
\State $append(elp_d, e)$
\If {$|elp_d| == mblk/\mathcal S(e)$} \Comment{Edge list packet is full.}
\State $mpi\_send(d, elp_d)$ \Comment{Send packet to collector on compute node $d.$}
\State $clear(elp_d)$
\EndIf
\EndFor
\For {$i \in [0:nb]$}
\State $mpi\_send(elp_{i}, i)$
\EndFor
\end{algorithmic}
\end{algorithm}

\begin{algorithm}
\caption{collect\_edges()}
\begin{algorithmic}[1]
\While{}
\If {$recv\_all\_packets()$}  \Comment{Check if all packets have been collected}
\State break
\EndIf
\State $mpi\_recv(elp)$ \Comment{MPI level details are omitted for clarity}
\State $append(el, elp)$
\EndWhile
\end{algorithmic}
\end{algorithm}

\subsubsection{Build CSR}
After the redistribute step each compute node has collected its edgelist. 
The final step requires conversion from edgelist to CSR representation. 
We use a parallel approach to make this step scalable. 
This problem was also considered in~\cite{DBLP:reference/parallel/Madduri11} but assuming OpenMP
language and for the main memory. Here, we consider a pthread implementation optimized 
for external memory.

The edgelist is again divided into $nc$ chunks. Each core scans its chunk (algorithm~\ref{alg:builddegv}). 
It maintains an associative map $degh$ (in memory) to keep track of the  degree of each node.  
Every time the size of associative map grows beyond a threshold, it updates
$degv,$ the degree vector of the graph nodes on that compute node. All updates to $degv$ vector are 
synchronized, using atomics, so as to avoid race condition. 

The offset vector $offv$ is then built sequentially as
$$offv[i] = offv[i-1] + degv[i] \forall i \in [1: n+1],$$ 
$offv[0] = 0.$ Finally, we build the $adjv$ (in parallel) to obtain the CSR representation.
Algorithm~\ref{alg:adjv} describes the steps and is similar to $build\_degv$ with the following differences:
one, the associative map is $adjvh$ instead of $degh$ and it stores the adjacent nodes instead of the degree, and
two, instead of adding the degree, we copy the adjacencies in the  $offv$ to its desired location.

The complexity of the CSR step is at most $O(b)$ random I/Os. 
The data structures $degh$ and $adjvh$ help alleviate the problem to some extent by aggregating writes 
to disk blocks. However, this amortization of random I/O cost becomes less and less effective with
large scale graph size. 

\subsubsection{Alternative Redistribute + CSR}
\label{sec:alt_csr}
\begin{figure}
\centering
\includegraphics[width=2.00in]{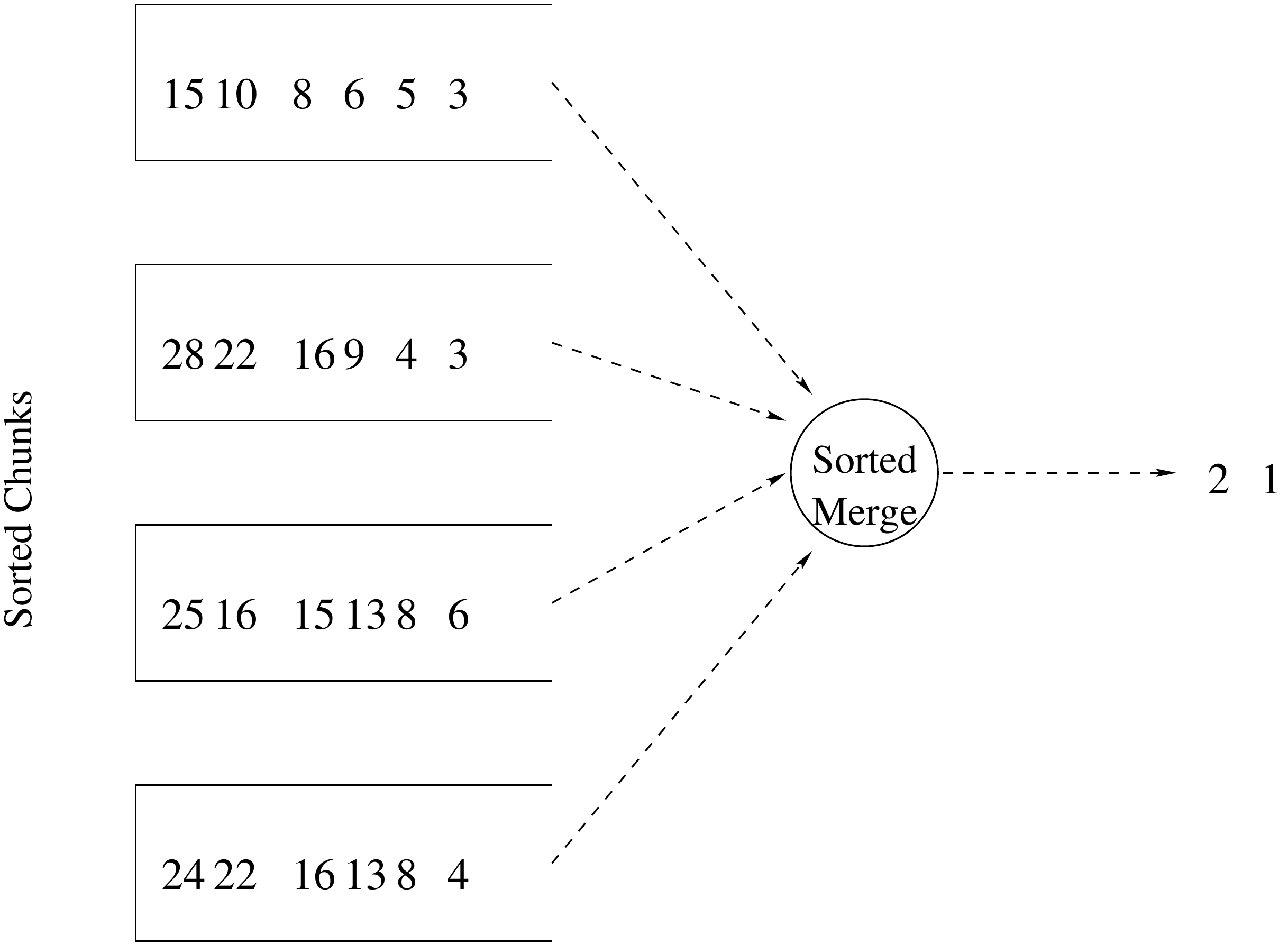}
\caption{Sorted merge operation applied at both send and receive  end of the redistribute step to generate an ordered sequence of edges. 
This help parallelize the workload and significantly reduce the complexity of the build CSR step. \label{fig:sorted_merge}}
\end{figure}

An alternative and more efficient scheme for redistribution and build csr using sorted merge operation is possible. Sorted 
merge operation produces a new sorted array  by combining a collection of existing sorted arrays. This is illustrated 
in  figure~\ref{fig:sorted_merge} . Unlike the previous CSR scheme whose performance is subject to ordering of edges, this scheme is 
guaranteed to require no more than $O(B/C_e)$ sequential I/Os. However, this scheme is not implemented in this paper. 

After the relabeling, we sort the edgelist chunks  again based up on the relabeled source field. These sorted
edgelists are merged (using sorted merge style operation) to produce a sorted sequence of edges. This sorted sequence
is redistributed to all compute nodes as described earlier in section~\ref{sec:redise}. 
The collector thread, collects all the received edges and stores them on external 
memory. It further  performs a similar operation on the received edges. Doing so makes all the edges (owned by the compute node) sorted based up on the source field. 
Once the edges are thus sorted, building csr representation becomes trivial and can be accomplished using the algorithm~\ref{alg:buildcsr}.
Hence the total I/O cost of building CSR representation using this approach would be $O(B/C_e)$ sequential reads. This can further improved to
$O(b/C_e)$ via simple parallization across the cores.

\begin{algorithm}
\caption{build\_degv(tid)}
\label{alg:builddegv}
\begin{algorithmic}[1]
\For {$k \in |elc|$}
\For {$(s,d) \in elc_k$}
\State $inc(adjv_h[s], 1)$
\If {$S(adjv_h[s]) == mmc$}
\State $degv[s] = degv[s] + |adjv_h[s]|$ \Comment{this is performed atomically}
\EndIf
\EndFor
\EndFor
\end{algorithmic}
\end{algorithm}

\begin{algorithm}
\caption{build\_edgev(tid)\label{alg:build_edgev}}
\label{alg:adjv}
\begin{algorithmic}[1]
\Require $deg[i] = 0$ for $i \in [0:B]$
\For {$k \in |elc|$}
\For {$(s,d) \in elc_k$}
\State $append(adjv_h[s], d)$
\If {$\mathcal S(adjv_h) == mmc$}
\Repeat
\State $do = degv[s]$
\State $dn = do + |adjv_h[s]|$
\Until{$CAS(adjv_h[s],do, dn)$}
\State $adjv[do:dn] = adjv_h[s]$ \Comment{copy all the adjacent nodes}
\State $delete(adjv_h[s])$
\EndIf
\EndFor
\EndFor
\end{algorithmic}
\end{algorithm}



\section{Experiments}
The experiments presented here were performed on Trestles cluster hosted at SDSC. 
Each compute node contains four sockets, each with a 8-core 2.4 GHz AMD Magny-Cours processor, for a total of 32 cores per node. The nodes have 64 GB of DDR3 RAM, with a theoretical memory bandwidth of 171 GB/s. The compute nodes are connected via QDR InfiniBand interconnect, fat tree topology, with each link capable of 8 GB/s (bidirectional). 


The experiments are  divided into three parts: single node experiments, strong scaling experiments, and weak scaling experiments.

\subsection{Single Node Experiments}
The  single node experiments study the performance of the operations without the interference of the communication complexity involved.
This allows us to project the limiting factor in scaling graph generation  and the maximum sized graph practically feasible on 
one compute node. 

\begin{figure}
\includegraphics[scale=0.3,angle=-90]{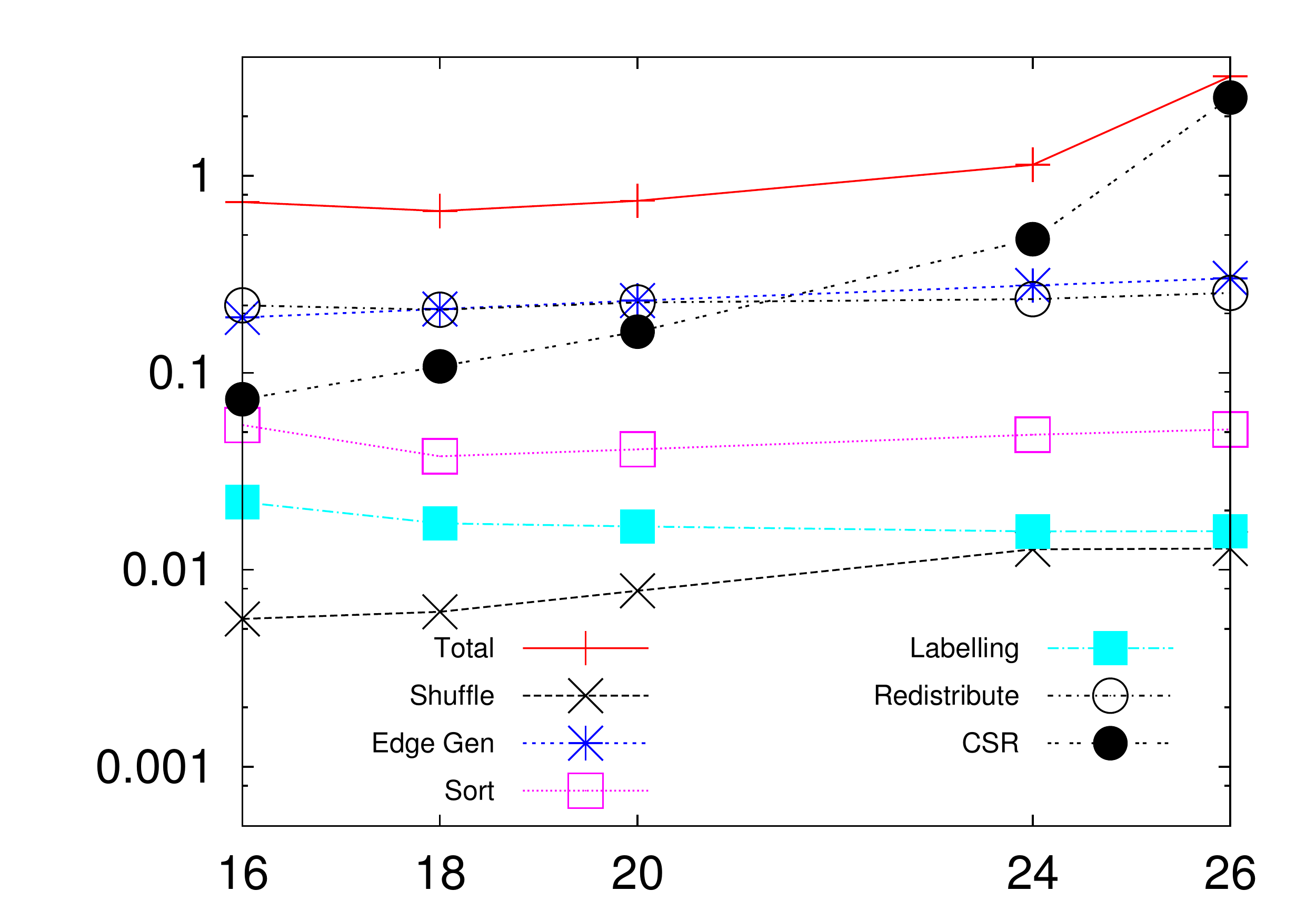}
\caption{Scalability on single compute node\label{fig:all_single_tt}}
\end{figure}

Figure~\ref{fig:all_single_tt} displays the compute time for each operation, normalized with respect to scale,
as the problem size is increased. 
The normalization was done in order to be able to compare/contrast the time across scales. 
Specifically, the actual time value was divided by $2^{s-16},$ where $s$ denotes the corresponding scale. 
Hence, if the time complexity grows linearly with respect to scale then we should expect an almost flat curve. 
We see this behavior for all operations including complex ones such as shuffle, labeling, and redistribute. 
CSR is an exception. It grows exponentially with scale, thereby, also increasing the total time. 
We argue this because our current implementation of CSR is not optimal. The version we presented in algorithm~\ref{alg:build_edgev} is 
guaranteed to scale linearly. 

Furthermore, the complexity analysis carried out in~\ref{sec:alt_csr} indicates linear scaling characteristics of the 
alternative build CSR implementation. 
Hence, from the experiment observations and the complexity analysis, we can claim that the dominant limiting factor is 
the size of the main memory. Even at that, except for shuffle, the rest of the algorithm can work with limited memory buffer.
In other words the operations can be performed with fixed amount of main memory buffer irrespective of the scale of the graph.
With this limitation we can generate a $2^{32}$ scale graph on a single compute node (assuming 64GB of main memory) within 
a few hours (2 - 3 hours). This implies that using 64 such compute nodes we can generate a $2^{38}$ sized graphs in less than 6 hours.
Furthermore, the limitation on the shuffle is artificial and can be lifted using external memory shuffle algorithm. Once this limitation
is removed the only true limitation would be the number of cores available in a system and IOPS and bandwidth of the external memory system. 

The current trends in multicore architecture and SSDs point towards increased number of cores on a single cpu as well
as IOPS and bandwidth on SSDs.  This implies that with our algorithm, in near future $2^{34}$ sized graph will also 
become feasible on a single compute node.  Hence, large scale graph generation which currently is possible only on the 
largest supercomputers in the world would be feasible on modest computing platform in near future.


\subsection{Strong Scaling}
\begin{figure}
\includegraphics[scale=0.3,angle=-90]{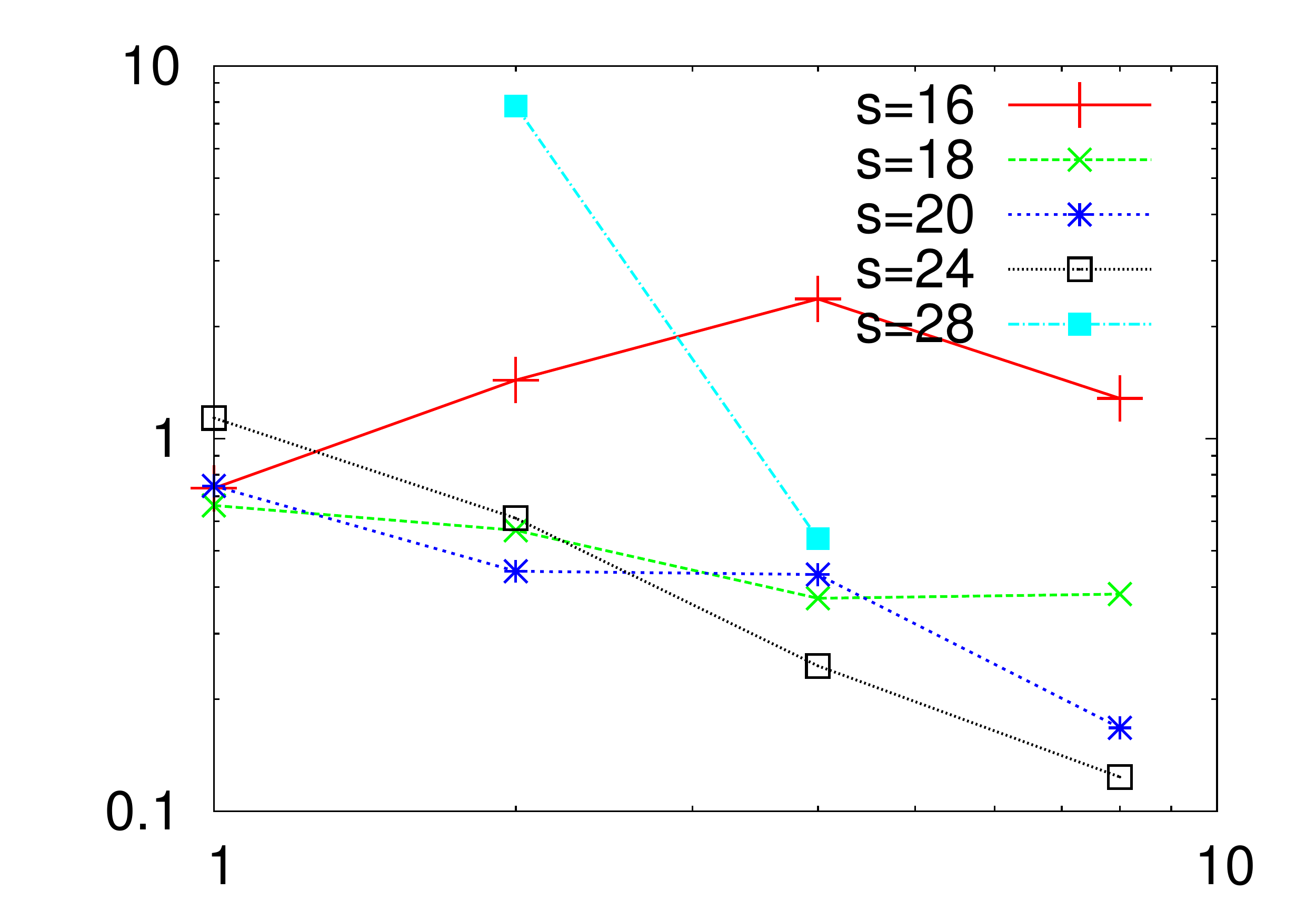}
\caption{Total generation time with increasing number of compute nodes for various scale graphs.\label{fig:tt}}
\end{figure}

\begin{figure}
\includegraphics[width=4.00in,angle=-90]{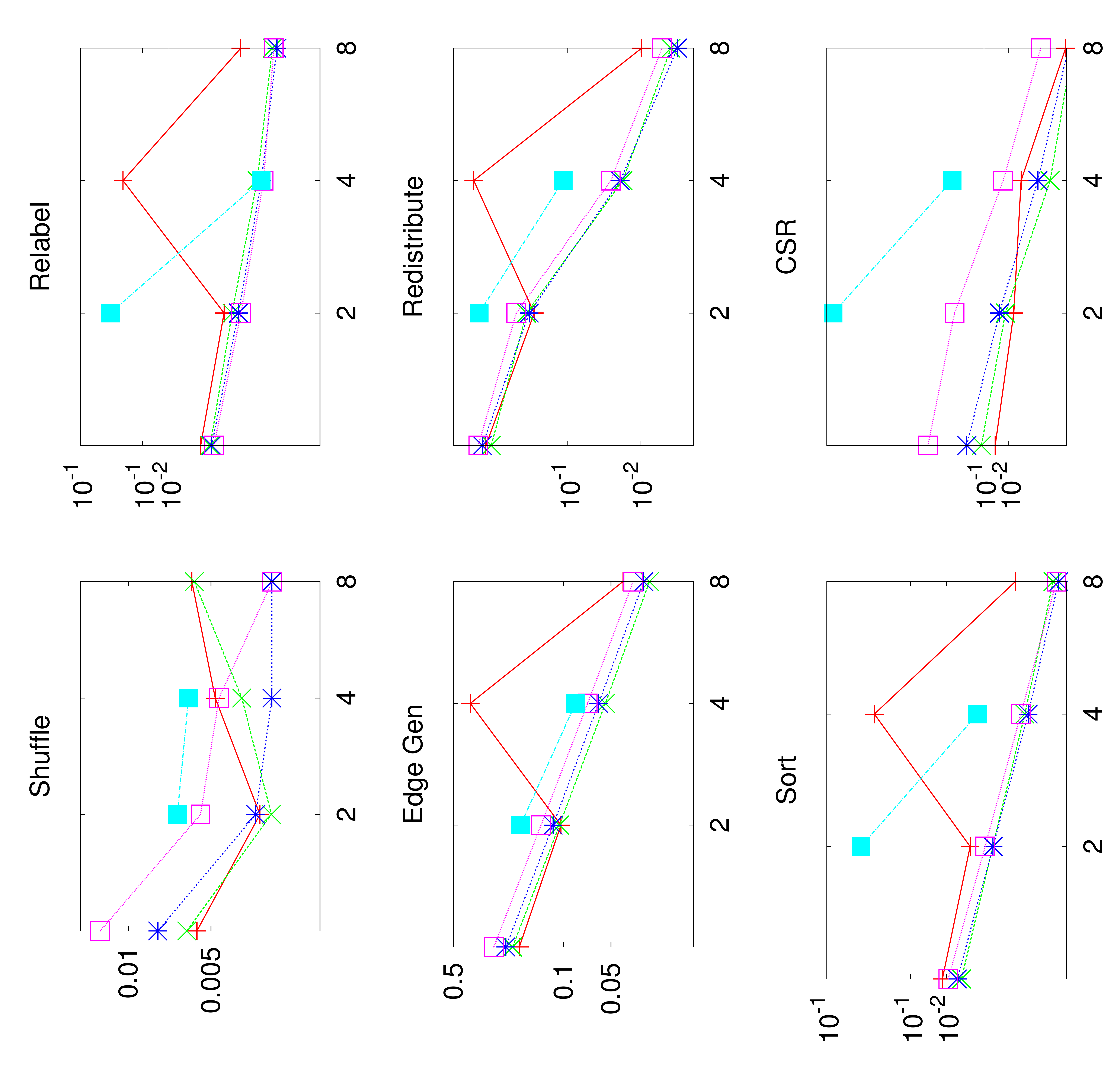}
\caption{Strong scaling characteristics of operations~\label{fig:all_tt}}
\end{figure}

We conduct experiments with increasing number of compute nodes for fixed problem sizes.
Figures~\ref{fig:tt} shows the total generation time while plots  in figures~\ref{fig:all_tt} show time for 
individual operations for scale 16, 18, 20, 24, 28 problems sizes as we increase the number of compute nodes
from 1 to 8. 
In order to compare the performance and to bring out the trend clearly,  the y-axis values (time taken for a operation) is scaled with 
respect to scale 16 problem size. Specifically, the y-axis values for problem at  scale $s$ is divided by $2^{s-16}$. This make sure 
that values are within the same range and we can compare the performance and trends across problems of varying scale. 

We see that the total time decreases linearly with respect to number of compute nodes. That is our approach 
exhibit linearly strong scaling. However, this scalability is limited by the problem scale. For example,
scale 16 graph hits the limit at  2 compute nodes while scale 18 graph scalability tapers off at 4 compute node.
On the other hand scale 28 graphs shows excellent scalability when the number of compute nodes are increased from 2 to 4. 
Data points for scale 28 graph for 8 compute node is missing because we ran out of our quota to run experiments. 

The scaling of the individual operations is shown in figure~\ref{fig:all_tt}. We see that edge generation, sort, and, csr 
scale linearly. The shuffle and  relabel show sublinear scalability. The redistribute operation has  unique behavior. 
For the sizes considered here it exhibits strong scaling. However, at large scale it also exhibits sublinear scalability 
as shown later in section~\ref{sec:weak_scale}. This is due to well known skewness  in degree distribution of social network 
graphs.

\subsection{Weak Scaling}
\label{sec:weak_scale}
Except `redistribute' and `relabel', all the operators in the generator are embarrassingly parallel and,
therefore, exhibit perfect weak scaling, i.e., the computation time of the operation remains constant upon increasing the
problem size if the number of compute nodes in increased proportionally.

\begin{figure}[!ht]
\includegraphics[width=2.6in,angle=-90]{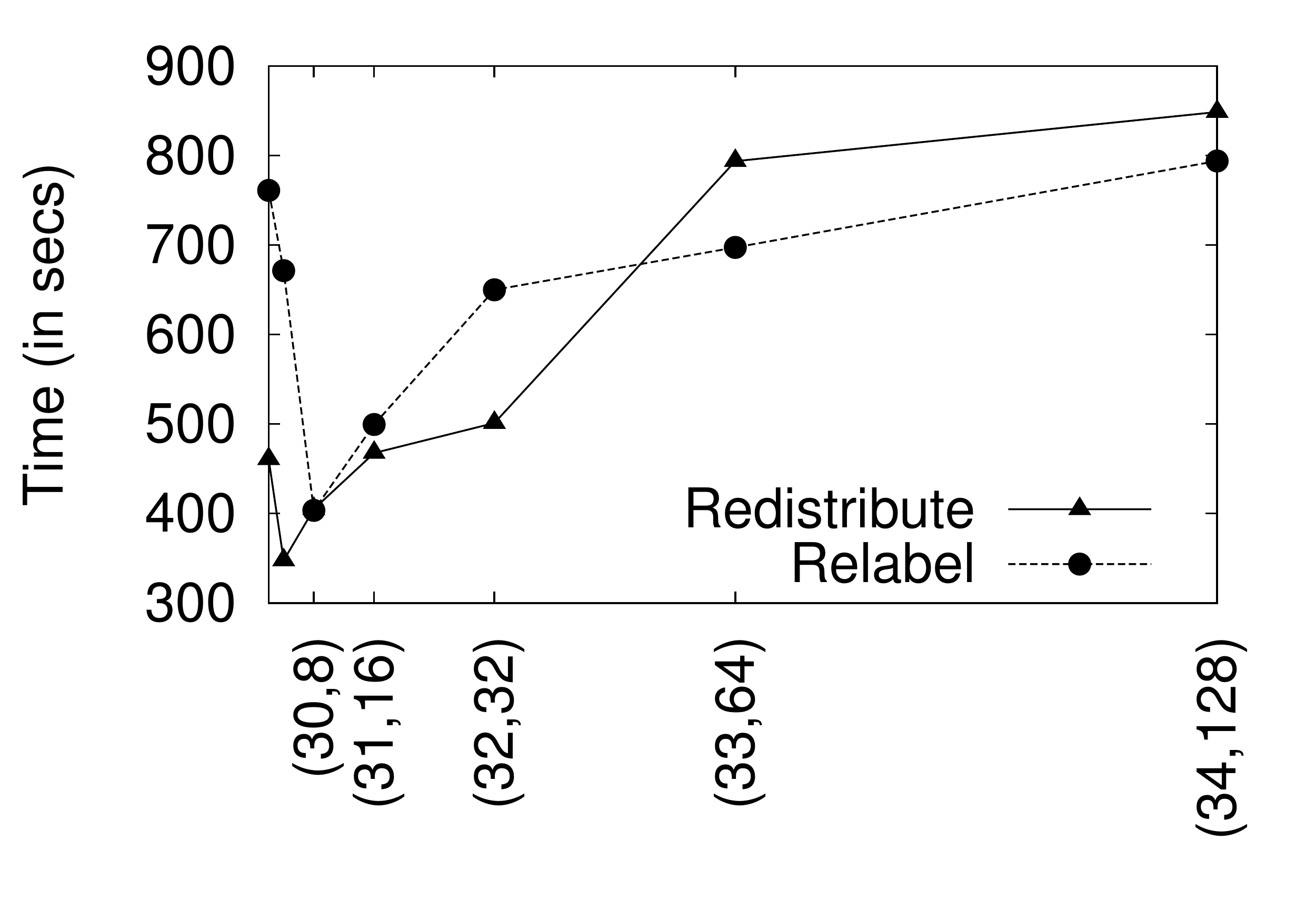}
\caption{Weak scaling characteristics of labeling and redistribute operators.\label{fig:weak_scaling}}
\end{figure}

Redistribution and relabeling do not exhibit perfect (weak) scalability. Figure~\ref{fig:weak_scaling} displays 
the scalability of these two operations as we increase both the problem size and the number of compute
nodes proportionally, starting with scale 29 graph and 4 compute nodes. Peformance at scale 28 with 2 compute nodes is abnormally high
due to skewness in the datasets. As the number of nodes is increased the skewness affects are mitigated.
The problem size was increased up to scale 34 while
the number of compute nodes were increased proportionally. So we ran the experiments for graphs of scales 29 to 34 using 8, 16, 32, 64, 
and, 128 compute nodes. The $x$-axis shows the problem size and number of compute nodes tuples, $(s,nb).$
The $y$-axis shows the time to perform labeling and redistribute.

We see that the graphs are not constant and grow sub-linearly with the problem size. In case of relabeling this is expected since each 
compute node scans the entire permutation vector $P$ which grows linearly with time. In case of redistribute operator, each compute node
scans the edges that belong to its partition. Hence, one would expect the time taken by it to be constant. Unfortunately, the time
taken increases with the increase in problem size. This is because the degree distribution of R-MAT graph is a skewed distribution.
As we weak scale the problem there are some compute nodes have unfairly high number of edges they own, thereby increasing
time complexity.

We remark that, this aspect does not in any way limit the scalability of our approach in terms of size.
The use of the external memory allows for sufficient room to accommodate for the skewness in the edge distribution.

